\documentclass{article}
\def\bfkappa{\mathord{\mbox{\boldmath $\kappa$}}}

\begin{document}
\begin{center}
{\bf SOME APPLICATIONS OF FRACTIONAL EQUATIONS}
\end{center}

\begin{center}
{\bf H. Weitzner$^{1)}$   and G.M. Zaslavsky$^{1,2)}$ \\
$^{1)}$Courant Institute of Mathematical Sciences,
New York University \\
251 Mercer St., New York, NY 10012, USA \\
$^{2)}$Department of Physics, New York University,
2-4 Washington Place, New York, NY 10003, USA
}
\end{center}

\begin{abstract}
We present two observations related to the
application of linear (LFE) and nonlinear
fractional equations (NFE). First, we give the comparison
and estimates of the role of the fractional derivative term to the
normal
diffusion term in a LFE.  The transition of the solution
from normal to anomalous transport is demonstrated
and the dominant role of the
power tails in the long time asymptotics is shown. Second,
wave propagation or kinetics in a nonlinear media with
fractal properties is considered.
A corresponding fractional generalization of the
Ginzburg-Landau and nonlinear Schr\"{o}dinger equations is proposed.
\end{abstract}
\newpage

\section{Introduction}

We call a fractional equation (FE) a differential equation that
contains fractional derivatives or integrals.
The awareness of the importance of this
type of equation has grown
continually in the last decade. Numerous
applications have become apparent:
wave propagation in a complex or porous media
\cite{ref:1}-\cite{ref:3};
random walks with a memory and flights
\cite{ref:4}-\cite{ref:6};
kinetic theories of systems with chaotic
\cite{ref:7}-\cite{ref:9}
and pseudochaotic
\cite{ref:10}
dynamics; and others (see
\cite{ref:9}
and collections of papers in
\cite{ref:11}).
This new type of problem has increased rapidly in
areas in which the fractal
features of a process or the medium impose the
necessity of using
non-traditional tools  in
``regular'' smooth physical equations.
Exploitation of the  fractional calculus for FE provides not only new types of
mathematical constructions, but also new physical features of the
described phenomenon
\cite{ref:4},\cite{ref:5}.
While the linear FE has attracted fairly broad research activity, the
study of nonlinear FE is at its very beginning
\cite{ref:11}-\cite{ref:14}.

This paper consists of two parts on the FE properties. The first part is
related to a linear FE (LFE) with the second (diffusional) derivative
and the fractional derivative of order $\alpha$. A goal of this part is to
present a solution to the FE and compare roles of different parts of the
FE.
The second part is related to a nonlinear FE (NLFE) with cubic
nonlinearity. We discuss some features of these equations,
and possible applications. For a broad discussion of
the fractal features of different media where waves propagate,
see the elegant review
\cite{ref:3}).
In particular, a turbulent media of this type  may be considered.
We will show that such media lead to NLFEs in a natural way
after simple change of the dispersion law.

\section{The interaction of normalized and anomalous transport}

The interest in and relevance of kinetic equations with fractional
derivatives is a natural consequence of the realization of the importance
of non-Gaussian properties of the statistics of many dynamical systems.
There is already a substantial literature studying such equations in one
or more space dimensions. In many cases of physical interest it would be reasonable
to assume that both Gaussian and anomalous processes would play a role.
One typically associates anomalous processes with algebraically decreasing
tails of a probability distribution function (PDF), while the bulk of the
PDF is expected to be mostly Gaussian in chracter. In this note we
explore the interaction of Gaussian and anomalous dynamics by means of a
simple model kinetic equation in one space dimension for the PDF.
More space dimensions would greatly complicate the analysis, but could
readily be carried out.

We consider the kinetic equation with fractional derivatives
\begin{equation}
{\partial P (x,t) \over \partial t^{\prime} } =
  {\partial^2 P\over \partial x^{\prime 2} }   +
\epsilon
{\partial^{\alpha} P \over \partial |x^{\prime}|^{\alpha} } \ ,
\ \ \ \ 1 < \alpha < 2 \label{eq:1}
\end{equation}
where $\epsilon$ is a constant, presumed small, but not necessarily so.
The Riesz derivatives
$\displaystyle{
{\partial^{\alpha} P \over \partial |x^{\prime}|^{\alpha} } }$
is easily defined for our purposes in Fourier transformed space, so that
the Fourier transform of
$\displaystyle{
{\partial^{\alpha} P \over \partial |x^{\prime}|^{\alpha} } }$
is $-|k|^{\alpha} \tilde{P} (k,t)$ where
$\tilde{P} (k,t)$ is the Fourier transform of $P (x^{\prime} ,t)$.
We have not taken a fractional derivative with respect to time. We could
do so, but this would be yet another complication in the analysis.
We see that the Fourier transform of the right hand side of (\ref{eq:1}) is
$-(k^2 + \epsilon |k|^{\alpha} ) \tilde{P} (,t)$. Thus, for large
wavenumber and short wavelength the system
exhibits normal, Gaussian transport,
while for small wavenumber and large wavelength, the system is anomalous.
The two phenomena are equal for wavenumber $k_T$
\begin{equation}
k_T = \epsilon^{{1\over 2-\alpha}} . \label{eq:2}
\end{equation}
Thus (\ref{eq:1}) provides a simple model which
mixes both normal and anomalous
transport.

It is convenient to change the independent variables from
$x^{\prime} ,t^{\prime}$ to $x,t$ with the definitions
\begin{equation}
x = \epsilon^{{1\over 2-\alpha}} x^{\prime} \label{eq:3}
\end{equation}
\begin{equation}
t = \epsilon^{{2\over 2-\alpha}} t^{\prime} \label{eq:4}
\end{equation}
so that
\begin{equation}
{\partial P (x,t) \over \partial t } =
{\partial^2 P \over \partial x^2 } +
{\partial^{\alpha} P \over \partial |x|^{\alpha} } \ .
\label{eq:5}
\end{equation}
The coordinate $(x^{\prime} ,t^{\prime} )$ are scaled to the Gaussian
transport scale, while $(x,t)$ are of a scale in which Gaussian and
anomalous transport are comparable. We may easily represent a solution of
(\ref{eq:5}) which also satisfies
\begin{equation}
P (x,0 ) = \delta (x) \label{eq:6}
\end{equation}
and as a consequence
\begin{equation}
\int_{-\infty}^{\infty} P (x,t) dx = 1 \ , \label{eq:7}
\end{equation}
as a Fourier integral:
\begin{equation}
P (x,t) = {1\over 2\pi} \int_{-\infty}^{\infty} dk
\exp [-t(k^2 + |k|^{\alpha} ) - ikx ] . \label{eq:8}
\end{equation}
We compare (\ref{eq:8}) with the solution
of the equation with only anomalous
transport
\begin{equation}
{\partial Q(x,t) \over \partial t} =
{\partial^{\alpha} Q(x,t) \over \partial |x|^{\alpha} } , \label{eq:9}
\end{equation}
where $Q$ also satisfies (\ref{eq:6}) and (\ref{eq:7}), so that
\begin{equation}
Q(x,t) = {1\over 2\pi} \int_{-\infty}^{\infty} dk
\exp (-t|k|^{\alpha} - ikx ) . \label{eq:10}
\end{equation}
The properties of $Q(x,t)$ have been well-explored in the literature. We
need such results for comparison with the properties of
$P (x,t)$.

We develop numerous series expansions of
(\ref{eq:8}) and (\ref{eq:10}), some convergent and
some asymptotic. When we know the behavior of $P (x,t)$ and $Q(x,t)$
for $|x| $ large, we shall also evaluate one spatial moment of
$P (x,t)$ and $Q(x,t)$. With this information in hand we can then
provide a description of the interaction of normal and anomalous transport.
We start with (\ref{eq:8}), which we rewrite as
\begin{equation}
P (x,t) = {{\rm Re} \over \pi} \int_0^{\infty} dk
\exp [-t (k^2 + k^{\alpha} ) -ikx ] \label{eq:11}
\end{equation}
We observe that
\begin{equation}
e^{-tk^{\alpha} } = \sum_{n-0}^N (-1)^n t^n
{k^{\alpha n} \over n!} + R_{N+1} (k,t) , \label{eq:12}
\end{equation}
where
\begin{equation}
|R_{N+1} (k,t)| \leq t^{N+1}
{k^{\alpha (N+1)}\over (N+1)!} . \label{eq:13}
\end{equation}
Since
\[
\left| \int_0^{\infty} dk
\exp (-tk^2 - ikx ) \right| \leq {t^{(N+1)} \over (N+1)!}
\int_0^{\infty} dk \ k^{\alpha (N+1)} e^{-tk^2}
\]
it is easy to conclude that the integral of the error tends to zero
uniformly for all $x$ and $0 < \eta \leq t \leq T$ as
$N \rightarrow\infty$, so that
\begin{equation}
P (x,t) = {{\rm Re} \over \pi} \sum_{n=0}^{\infty} (-1)^n {t^n \over n!}
\int_0^{\infty} dk \ k^{\alpha n} \exp (-tk^2 - ikx ) , \label{eq:14}
\end{equation}
or
\begin{equation}
P (x,t) = {1\over 2\sqrt{\pi t} }
e^{- {x^2 \over 4t}} + {\rm Re} {e^{-x^2 /8t}\over \pi \sqrt{t} }
\sum_{n=1}^{\infty} (-1)^n
{t^{n(1 - {\alpha\over 2} )} \Gamma (\alpha n+1) \over
  2^{{\alpha n+1 \over 2}} \Gamma (n+1) }
D_{-\alpha n-1} \left( {ix \over \sqrt{2t} } \right) , \label{eq:15}
\end{equation}
where $D_{\nu} (z)$ is the Weber function of order $\nu$.
Again (\ref{eq:15})
converges uniformly for all $x$ in $0 < \eta \leq t \leq T$. We see
that the first term in (\ref{eq:15}) exhibits
Gaussian diffusion and also has unit
integral. Hence, the remaining terms, the anomalous transport effects,
have zero mean. For $|x|$ large and $n$ fixed we may employ the asymptotic
expansion of the Weber function of
large argument, or directly from (\ref{eq:15}) we obtain the
asymptotic expansion for large $|x|$:
\[
{1\over \pi} {\rm Re} (-1)^n {t^n \over n!} \int_0^{\infty} dk \
k^{\alpha n} \exp (-tk^2 - ikx )
\]
\begin{equation}
=
{(-1)^{n+1} \over \pi} \
{t^b \over n! |x|^{\alpha n+1} } \sin \left( {\alpha \pi n\over 2} \right)
\left\{ \sum_{m=0}^M \left( {t\over x^2} \right)^m
{\Gamma (\alpha n + 2m+1) \over m!} + O \left( {t\over x^2 } \right)^{M+1}
\right\} \label{eq:16}
\end{equation}
The series (\ref{eq:16}) is clearly asymptotic and
shows the power tail of the PDF, with
the leading term from $n=1$ being
\begin{equation}
{(t/|x|^{\alpha} ) \over \pi |x| } \Gamma (\alpha +1) \sin
{\pi \alpha \over 2} \ . \label{eq:17}
\end{equation}

We may obtain another convergent expansion by use of the power series
\begin{equation}
e^{-ikx} = \sum_{n=0}^N (-i)^n {k^n x^n \over n!} + R_{N+1} (k,x) ,
\label{eq:18}
\end{equation}
where
\begin{equation}
|R_{N+1} (k,x) | \leq k^{N+1} {|x|^{N+1} \over (N+1)!} \label{eq:19}
\end{equation}
and thus
\begin{equation}
P (x,t) = \sum_{n=0}^{\infty} (-1)^n x^{2n} C_n (t) , \label{eq:20}
\end{equation}
where
\begin{equation}
C_n (t) = {1\over (2n)! \pi } \int_0^{\infty} dk  \ k^{2N}
\exp -t(k^2 + k^{\alpha} ) . \label{eq:21}
\end{equation}
Although $C_n (t)$ does ot appear to be one of the usual special functions
it is easy to obtain an asymptotic expansion for $t$ large and convergent
expansion for $t$ small. In particular for $t$ small but bounded away
from zero:
\begin{equation}
C_n (t) = {t^{-n- 1/2} \over (2n)! 2\pi } \sum_{n=0}^{\infty}
{(-1)^m \over m!} t^{m(1-\alpha /2) }
\Gamma \left( n + {\alpha m \over 2} + {1\over 2} \right) \label{eq:22}
\end{equation}
and for $t$ large
\begin{equation}
C_n (t) = {t^{-(2n+1) /\alpha } \over (2n)! 2\pi}
\sum_{m=0}^{\infty} {(-1)^m \over m!} t^{m(1 - {2\over \alpha} ) }
\Gamma \left( {2(n+m) +1 \over 2} \right) . \label{eq:23}
\end{equation}

We see from (\ref{eq:14}) and
(\ref{eq:17}) that for $|x|$ large $P(x,t) \sim ({\rm const.})
|x|^{-\alpha -1}$. Thus, moments of order $\geq \alpha$ do not exist. We
examine one, simple relevant moment
\begin{equation}
M = \int_{-\infty}^{\infty} |x| dx \ P (x,t) \label{eq:24}
\end{equation}
and it follows easily that
\begin{equation}
M = {2\over \pi} \int_0^{\infty} dx \ {\rm Re} \int_0^{\infty} dk \
\exp [ -t(k^2 + k^{\alpha} ) ] {d\over dk} \sin kx . \label{eq:25}
\end{equation}
After integration by parts we obtain
\begin{equation}
M = {2t \over \pi} \int_0^{\infty} dx \ {\rm Im} \int_0^{\infty} dk
(2k + \alpha k^{\alpha -1} ) \exp [-t(k^2 + k^{\alpha} )] e^{ikx}
\label{eq:26}
\end{equation}
If we move the path of $x$ integration into the upper half-plane we find
\begin{equation}
M = {2t \over \pi} \int_0^{\infty} dk (2+\alpha k^{\alpha -2} )
\exp - t(k^2 + k^{\alpha} ) \label{eq:27}
\end{equation}
With the same type of expansion as before we find an expansion
convergent in $0 < \eta \leq t \leq T$
\begin{equation}
M = {t^{1/2} \over \pi} \sum_{m=0}^{\infty} (-1)^{m+1}
{t^{m(1-\alpha /2)} \over m!}
\Gamma \left( {m\alpha -1 \over 2} \right) \label{eq:28}
\end{equation}
The leading order term $m = 0$ is just $2(t/\pi )^{1/2}$, exactly what
one would obtain from a Gaussian. The remaining terms are the
corrections from anomalous transport. For $t$ large we have the
asymptotic expansion
\begin{equation}
M = {2t^{1/\alpha} \over \pi} \sum_{m=0}^{\infty} (-1)^m
t^{-m({2\over \alpha} -1)}
{(- {1\over \alpha} ) \Gamma ({2m-1 \over \alpha} ) \over m!}
\label{eq:29}
\end{equation}

We now turn to $Q(x,t)$ so that we may assess the significance of the
preceding results. We start from (\ref{eq:10}) and we find easily the
convergent expansion
\begin{equation}
Q = {1\over \pi \alpha} \sum_{n=0}^{\infty} (-1)^n t^{-1/2}
[x/t^{1/\alpha } ]^{2n}
{\Gamma ({2n+1 \over \alpha } ) \over \Gamma (2n+1)} \label{eq:30}
\end{equation}
while for $|x|$ large there is the asymptotic expansion
\begin{equation}
Q \sim {1\over \pi |x| } \sum_{n=1}^{\infty} (-1)^n
(t/|x|^{\alpha} )^n
{\Gamma (n\alpha +1) \over \Gamma (n+1)} \sin
{n\pi\alpha\over 2} . \label{eq:31}
\end{equation}
Finally, we define $M_Q$ as
\begin{equation}
M_Q = \int_{-\infty}^{\infty} Q(x,t) |x| dx \label{eq:32}
\end{equation}
so that
\begin{equation}
M_Q = {2\over \pi} \int_0^{\infty} xdx \ {\rm Re} \int_0^{\infty}
dk \exp (-k^{\alpha} t) {d\over dk} \sin kx . \label{eq:33}
\end{equation}
We follow the same procedure as for $M$ and we obtain
\begin{equation}
M_Q = {2\over \pi} t^{1/\alpha} \Gamma (1-1/\alpha ) . \label{eq:34}
\end{equation}

We are now prepared to compare results and we start with the simplest
comparison, $M$ and $M_Q$. The result
(\ref{eq:34}) for $\nu$ is valid for all
times, however $M (t)$ is very different from
(\ref{eq:34}) for $t \sim 1$. For
$t$ large (\ref{eq:29}) shows
\begin{equation}
M \sim {2\over\pi} \left[ t^{1/\alpha} \Gamma \left( 1 - {1\over \alpha}
\right) + t^{1 - 1/\alpha} \Gamma \left( 1 + {1\over \alpha} \right) +
O (t^{2-{3\over \alpha} } ) \right] . \label{eq:35}
\end{equation}
Thus, in leading order and for $1 < \alpha < 2$, $M \sim M_Q$.
However, the difference between $M$ and $M_Q$ is not small unless
$t^{1/\alpha} \gg t^{1 - {1\over \alpha} }$. For $\alpha$ not far from one,
this relation holds, but as $\alpha $ approaches 2, one requires
increasing larger values of $t$ in order that $t^{1/\alpha}$ dominate
$t^{1-{1\over \alpha}}$. It is clear that the expansion (\ref{eq:29})
fails at $\alpha = 2$, and must then be very poor for
small values of $2-\alpha$.
An examination of this moment indicates that for $t$
sufficiently large the anomalous
transport is the limiting form for the case with both Gaussian and
anomalous transport, however there may be significant corrections.

When we compare $P (x,t)$ and $Q(x,t)$ the situation is somewhat
more complex. For $t$ and $x \stackrel{<}{\sim} 1$, so that both Gaussian
and normal transport may occur we may compare
(\ref{eq:15}),(\ref{eq:16}) with (\ref{eq:30}),(\ref{eq:31}).
When $|x| \sim 1$, $P $ and $Q$ are substantially different; however
for $t \sim 1$, $|x| > 1$, one may use (\ref{eq:16}),
and then the leading order terms in
(\ref{eq:16}) match (\ref{eq:3}) although there are
nontrivial corrections of order
$t/x^2$. When $t> 1$ but $|x| \sim 1$, we may compare
(\ref{eq:21}) and (\ref{eq:23})
with (\ref{eq:30}). Again the leading order terms agree but there are
nontrivial corrections of order $t^{-({2\over\alpha} -1)}$. Finally, when
both $x$ and $t$ are large, we must decide whether
$|x|^{\alpha} /t$ is large or not. If $|x|^{\alpha} /t \sim 1$
after a little effort we see that we may use
(\ref{eq:20}) and (\ref{eq:23}) for $P (x,t)$,
which again matches $Q(x,t)$ in (\ref{eq:30}),
but with corrections. Finally if $|x|$ and $t$ are both large, as is
$|x|^{\alpha} /t$, then we may conclude that
(\ref{eq:14}) and (\ref{eq:16}) apply for $P (x,t)$ which
approximates $Q(x,t)$ as given by (\ref{eq:31}). We see that broadly the
anomalous transport
finally dominates the Gaussian, or normal, transport,
although it may take a long time. From another point of view,
unless $\epsilon$ is extremely small,
the anomalous transport description is generally more
relevant than the normal transport, although there may be significant
corrections to the anomalous transport results.

\section{Fractional generalization of Ginzburg-Landau equation and
nonlinear Schr\"{o}dinger equation}

Let us recall the appearance of the nonlinear parabolic equation
(see for example
\cite{ref:15}).
The simplest way is to consider a symmetric dispersion law
$\omega = \omega (k)$ for wave propagation in some media, and to represent
the wave vector $\bf k$ in the form
\begin{equation}
{\bf k} = {\bf k}_0 + {\bfkappa} = {\bf k}_0 + {\bfkappa}_{\parallel}
+ {\bfkappa}_{\perp}
\label{eq:36}
\end{equation}
where ${\bf k}_0$ is the unperturbed wave vector and subscripts
$(\parallel ,\perp )$ are taken respectively to the direction
${\bf k}_0$. Considering $\kappa \ll k_0$ we have
\[
\omega (k) = \omega (|{\bf k}_0 + {\bfkappa}|) \approx \omega (k_0 )
\]
\begin{equation}
+ \ c(|{\bf k}_0 + {\bfkappa}| - k_0 )
  \approx \omega (k_0 ) + c\kappa_{\parallel} +
{c\over 2k_0} \kappa_{\perp}^2
    \label{eq:37}
\end{equation}
where $c = \partial\omega /\partial k_0$. The expression
(\ref{eq:37}) corresponds to the linear parabolic equation
\begin{equation}
- i {\partial Z \over \partial t} = ic {\partial Z \over \partial x}
  + {c\over 2k_0 } \Delta Z
    \label{eq:38}
\end{equation}
with respect to a field $Z = Z (x,{\bf r},t)$, $x$ along the ${\bf k}_0$,
and the operator correspondence:
\begin{equation}
\nu \equiv \omega (k) - \omega (k_0 ) = i {\partial\over\partial t} ,
\ \ \
\kappa_{\parallel} = -  i {\partial\over\partial x} , \ \ \
{\bfkappa}_{\perp} = -i {\partial\over\partial {\bf r} } , \ \ \
{\bf r } = (y,z)
\label{eq:39}
\end{equation}
A generalization to a nonlinear case can be carried out in analogy with
(\ref{eq:37}) through a nonlinear dispersion law depending on the wave
amplitude:
\[
\omega  =  \omega (k,|Z|^2 ) \approx\omega (k,0) + g|Z|^2
\]
\begin{equation}
  =  \omega (|{\bf k}_0 + {\bfkappa}| ,0) + g|Z|^2
  \label{eq:40}
\end{equation}
with some constant $g = \partial\omega (k, |Z|^2 )/\partial |Z|^2$ at
$|Z|^2 = 0$. In analogy with (\ref{eq:38}), the nonlinear parabolic
equation takes the form
\begin{equation}
- i {\partial Z \over \partial t}
= ic {\partial Z \over \partial x}
  + {c \over 2k_0} \Delta Z - g|Z|^2 Z
  \label{eq:41}
\end{equation}
This equation is also known as the nonlinear Schr\"{o}dinger equation
in which any of the coefficients may be complex. Indeed, for a
traveling wave $Z = Z(x-ct, {\bf r})$ we have
\begin{equation}
- ic {\partial Z \over \partial \xi}
  = {c \over 2k_0} \Delta Z - g|Z|^2
  \label{eq:42}
\end{equation}
with $\xi = x - ct$.

Wave propagation in a media with fractal propeties can be easily
generalized from the first step of writing dispersion law
(\ref{eq:37}). Namely, one can replace
(\ref{eq:37}) and (\ref{eq:40}) by the following:
\begin{equation}
\nu = \omega (k,|Z|^2 ) - \omega (k_0 ,0) = c\kappa_{\parallel} +
c_{\alpha} ({\bfkappa}_{\perp}^2 )^{\alpha /2} + g|Z|^2
  \label{eq:43}
\end{equation}
with a fractional value of $1 < \alpha < 2$ and new constant
$c_{\alpha}$. This replacement does not affect the nonlinear term which
appears for $\kappa_{\parallel} = \kappa_{\perp} = 0$
(see (\ref{eq:40})).

Using the connection between the differentiation operator and its
Fourier transform
\begin{equation}
(-\Delta )^{\alpha /2} \leftrightarrow ({\bfkappa}_{\perp}^2 )^{\alpha /2}
  \label{eq:44}
\end{equation}
we obtain the equation corresponding to
(\ref{eq:43}):
\begin{equation}
- i {\partial Z \over \partial t}
= ic {\partial Z \over \partial x}
  - {c \over 2k_0} (-\Delta )^{\alpha /2}  Z - g|Z|^2 Z
  \label{eq:45}
\end{equation}
or for travelling waves as in
(\ref{eq:42})
\begin{equation}
 ic {\partial Z \over \partial \xi}
  = {c \over 2k_0} (-\Delta )^{\alpha /2}  Z + g|Z|^2 Z
\label{eq:46}
\end{equation}
Let us comment on the physical structure of
(\ref{eq:46}).
The first term on the right-hand side
is related to the wave propagation in a media with fractal
properties. The fractal derivative may also appear
as a result of ray chaos
\cite{ref:16},
or of a superdiffusive wave propagation (see also the discussion in
\cite{ref:3},\cite{ref:9}
and corresponding references). The second term on the right-hand side of
(\ref{eq:45}),(\ref{eq:46}) corresponds to the wave interaction due to
the nonlinear properties of the media. Thus, (\ref{eq:46}) describes
fractional processes of self-focusing and related issues.

We may consider a one-dimensional simplification of (\ref{eq:46}), i.e.
\begin{equation}
 c {\partial Z \over \partial \xi}
= a {\partial^{\alpha} Z \over \partial |z|^{\alpha} }
  + g|Z|^2 Z
    \label{eq:47}
\end{equation}
with some generalized constants $c,a,g$ and then reduce
(\ref{eq:47}) to the
case
of a propagating wave solution
\begin{equation}
  Z = Z(z-c\xi ) \equiv Z (\eta ) .
    \label{eq:48}
\end{equation}
Then (\ref{eq:47}) takes the form of fractional generalization of the
Ginzburg-Landau equation (FGL):
\begin{equation}
 a {d^{\alpha} Z \over d |\eta |^{\alpha} }
+  {dZ \over d \eta  }
  + gZ^3 = 0 \label{eq:49}
\end{equation}
for real $Z(\eta )$. Now (\ref{eq:49}) differs from   the
fractional Burgers equation
\cite{ref:13},\cite{ref:14}
in the structure of the nonlinear term. Nevertheless,
an analysis similar
to \cite{ref:13},\cite{ref:14}
may be performed to obtain some estimates on the solution.

It is well known that the nonlinear term in the equations of
(\ref{eq:42}) type leads to a steepening of the solution and its
singularity. The steepening process may be stopped by a diffusional or
dispersional term, i.e. by a higher derivative term. A similar phenomenon
may appear for the fractional nonlinear equations
(\ref{eq:47}),(\ref{eq:49}). It has been shown in
\cite{ref:13}
that for the fractional Burgers equation there exists a critical value
$\alpha_c$ such that  solution is regular for all time
if $\alpha > \alpha_c$.

Next, we consider the symmetric solution $Z (\eta ) = Z(-\eta )$ of
(\ref{eq:48}) and assume $Z \in I\!\! R^2$. Let us multiply
(\ref{eq:49}) by $Z$ and integrate it under the assumption that
$Z$ is sufficiently small at infinity. We find
\begin{equation}
{\cal E} \equiv \int_{-\infty}^{\infty} (aZ
Z^{(\alpha /2)} + gZ^4 ) d\eta = 0
  \label{eq:50}
\end{equation}
where we use the notation
\begin{equation}
{d^{\alpha} Z \over d|\eta |^{\alpha} } \equiv Z^{(\alpha )}
\label{eq:51}
\end{equation}
The expression (\ref{eq:50}) gives a conservation law showing the
relative importance of the nonlinear and dispersion terms.
Considering $Z^{(\alpha )} \sim 1/\eta^{\gamma}$ we conclude that the
term with derivative prevails for small $\eta$ if
\begin{equation}
  \alpha > 2 \gamma .
  \label{eq:52}
    \end{equation}
For $2 > \alpha > 1$ the condition (\ref{eq:52})
shows that the solution is square
integrable at $\eta \rightarrow 0$.

Finally we observe that (\ref{eq:49}) may describe a new type of
fractional dynamic motion.
This topic will be considered elsewhere.

\section{Conclusion}

In this brief paper we have
made two remarks related to application of fractional
equations in physics. The first remark is related to the competition
between normal diffusion and diffusion induced by
fractional derivatives, as occur in
fractional kinetic theories. It is shown that for large times
the fractional derivative
term dominates the solution and leads to
power type tails.
The second remark is related to a new class of equations in which
fractional derivative terms are responsible for fractional dispersion.
In this case the asymptotics should be defined by a competition between
the fractional dispersion and nonlinear terms. We discuss the origin of the
fractional Ginzburg-Landau equation and fractional nonlinear
Schr\"{o}dinger equation.

\section{Acknowledgements}

H.W. was supported by the U.S. Department of Energy, Grant No.
DE-FG02-86ER53223. G.Z. was supported by the U.S. Navy Grants No.
N00014-96-1-00557 and No. N00014-97-1-0426, and by the U.S. Department
of Energy, Grant No. DE-FG02-92ER54184.

\end{document}